\documentclass[aps,prd,twocolumn,reprint,amsmath,amssymb,preprintnumbers,superscriptaddress,nobibnotes,nofootinbib,floatfix]{revtex4-1}

\usepackage{graphicx}
\usepackage{dcolumn}
\usepackage{bm}
\usepackage{color}
\usepackage{amssymb}
\usepackage{amsmath}
\usepackage[colorlinks=true,allcolors=Purple,pdfusetitle]{hyperref}
\usepackage{rotating}
\usepackage{multirow}
\usepackage{graphicx}
\usepackage[dvipsnames]{xcolor}
\usepackage{svg}
\usepackage{esdiff}
\definecolor{Green}{RGB}{0, 128, 0}
\newcommand{\orcid}[1]{\href{https://orcid.org/#1}{\includegraphics[width=10pt]{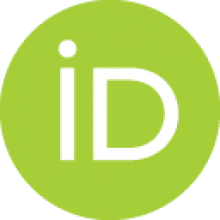}}}

\newcommand{\prlsection}[2]{{\it\textbf{#1}{#2}}---}
\makeatletter

\begin{document}

\preprint{FERMILAB-PUB-25-0596-T}
\preprint{KEK-QUP-2025-0020}
\preprint{KEK-TH-2752}

\title{Oscillation-independent probes of nonstandard neutrino interactions from supernovae}

\author{Angela R. Beatty}
\altaffiliation{\href{mailto:abeatty@sfsu.edu}{abeatty@sfsu.edu}}
\affiliation{Department of Physics and Astronomy, San Francisco State University, 1600 Holloway Ave, San
Francisco, 94132, California, USA}
\affiliation{
Department of Physics, University of California Berkeley, Berkeley, California 94720, USA}

\author{Anna M. Suliga\orcid{0000-0002-8354-012X}}
\altaffiliation{\href{mailto:a.suliga@nyu.edu}{a.suliga@nyu.edu}}
\affiliation{Center for Cosmology and Particle Physics, New York University, New York, NY 10003, USA}

\author{Volodymyr Takhistov \orcid{0000-0003-2647-3431}}  
\altaffiliation{\href{mailto:vtakhist@post.kek.jp}{vtakhist@post.kek.jp}}
\affiliation{International Center for Quantum-field Measurement Systems for Studies of the Universe and Particles (QUP, WPI),
High Energy Accelerator Research Organization (KEK), Oho 1-1, Tsukuba, Ibaraki 305-0801, Japan}
\affiliation{Theory Center, Institute of Particle and Nuclear Studies (IPNS), High Energy Accelerator Research Organization (KEK), Tsukuba, Ibaraki 305-0801, Japan
}
\affiliation{Graduate University for Advanced Studies (SOKENDAI), 
1-1 Oho, Tsukuba, Ibaraki 305-0801, Japan}
\affiliation{Kavli Institute for the Physics and Mathematics of the Universe (WPI), UTIAS, \\The University of Tokyo, Kashiwa, Chiba 277-8583, Japan}
 
\date{June 27, 2026}


\begin{abstract}
Extreme astrophysical environments provide unique laboratories for testing fundamental neutrino interactions. We present the first oscillation-independent astrophysical probe of nonstandard neutrino interactions (NSI), using coincident neutral-current signals across diverse detectors to break degeneracies that have long limited sensitivity reach.
Using self-consistent NSI supernova simulations and flavor-independent neutral-current scattering we show that anti-correlated signatures between JUNO liquid scintillator and dark matter detectors such as DARWIN/XLZD, ARGO, or RES-NOVA enable clear discrimination between NSI and flavor-conversion effects. For a Galactic supernova at Betelgeuse distance our approach enables an independent probe of neutrino-quark NSI couplings in parameter space that can reach and extend beyond current terrestrial limits. This multidetector approach enables breaking degeneracies in terrestrial searches and is broadly applicable to a wide range of upcoming experiments, establishing a new principle for testing fundamental interactions with astrophysical data.
\end{abstract}

\maketitle

\prlsection{Introduction}{.}
\label{sec:intro}
The detection of neutrinos from supernova (SN)~1987A~\cite{PhysRevLett.58.1490,Bionta:1987qt,ALEXEYEV1988209} established multi-messenger astronomy of transient events. While only a handful of $\bar\nu_e$ were observed, the subsequent discovery of neutrino masses through observation of neutrino oscillations in solar and atmospheric neutrino fluxes~\cite{Super-Kamiokande:1998kpq, SNO:2002tuh} demonstrated physics beyond the Standard Model (SM) in the neutrino sector.
New fundamental ``nonstandard neutrino interactions'' (NSI) with matter~\cite{Wolfenstein:1977ue, Valle:1987gv, Dev:2019qno} arise in many theories and can strongly affect neutrinos in core-collapse supernovae (CCSNe)~\cite{Nunokawa:1996tg,Lei:2019nma, Mikheyev:1985zog, Mansour:1997fi, Fogli:2002xj, Esteban-Pretel:2009jqw, Blennow:2008er, Stapleford:2016jgz, Farzan:2018gtr, Suliga:2020jfa, Cerdeno:2021cdz, Cerdeno:2023kqo, Dutta:2025rxh}. Terrestrial probes constrain neutrino-quark NSI primarily through their effects on the neutrino oscillations, \emph{e.g.}, Ref.~\cite{Super-Kamiokande:2022lyl} or coherent elastic nucleus-neutrino scattering~\cite{COHERENT:2017ipa, COHERENT:2020iec, XENON:2020gfr, CONUS:2021dwh, COHERENT:2025vuz}.

The extreme conditions of CCSN comprising a dense, hot and degenerate environment constitute an ideal laboratory for probing fundamental neutrino properties~\cite{Bruenn:2012mj, Takiwaki:2013cqa, OConnor:2018sti, Janka:2016fox, Mezzacappa:2020oyq, Burrows:2020qrp}. 
Neutrino-quark NSI can strongly affect CCSN emission during the pre-neutronization phase preceding the neutronization burst~\cite{Huang:2021enl}. Within $\lesssim 1$ kpc over 30 candidate progenitors that could undergo core-collapse in the coming decades have already been identified~\cite{Mukhopadhyay:2020ubs}. Among the closest is Betelgeuse ($\alpha$-Ori), with mass $16.5$-$19~M_{\odot}$ and distance $168^{+27}_{-15}$ pc ~\cite{Joyce:2020}. Detecting this early ``pre-SN'' neutrino signal is of particular interest, and a dedicated early-warning system has recently been implemented by Super-Kamiokande and KamLAND~\cite{KamLAND:2024uia}.

In this work we introduce the first astrophysical method to probe NSI independently of flavor conversions in CCSN using neutral-current (NC) coincidence signatures across pairs of detectors. Distinct flavor independent NC channels, exhibiting opposite responses in different experiments, produce anti-correlated signatures that 
break cross section and luminosity degeneracies limiting single-channel studies. We demonstrate this with self-consistent CCSN simulations that incorporate NSI effects on neutrino opacities in the proto-neutron star and propagate the resulting fluxes to predict observable signatures in terrestrial detectors.
 
\prlsection{Nonstandard neutrino interactions}{.}
\label{sec:NSI}
We consider vector NC neutrino NSI 
generated by a heavy mediator, described by the effective four-fermion interaction Lagrangian~\cite{Wolfenstein:1977ue, Dev:2019qno}
\begin{equation}
\mathcal{L}_{\rm NSI} = -2 \sqrt{2} G_\mathrm{F}  \epsilon_{\alpha\beta}^{fV} (\overline{\nu}_{\alpha L} \gamma^{\rho}\nu_{\beta L})(\overline{f}\gamma_{\rho}f
) \ ,
\end{equation}
where $G_\mathrm{F} = 1.1664 \times 10^{-5}$~GeV$^{-2}$ is the Fermi constant, $\alpha, \beta = e, \mu, \tau$ represent neutrino flavors, and $f = u, d$ are the quark fields. Here, we consider NSI interactions exclusively with first generation quarks. Further, we focus on flavor-conserving diagonal positive NSI, with $\alpha = \beta$, interacting with all neutrino flavors $\nu_e$, $\nu_{\mu}$ and $\nu_{\tau}$. We take the NSI couplings between neutrinos and quarks $\epsilon^{u (d)} = \epsilon_{ee}^{u(d)V} = \epsilon_{\mu\mu}^{u(d)V} = \epsilon_{\tau\tau}^{u(d)V}$. 

We implement NSI effects on interaction cross sections assuming that only one quark coupling is nonzero at a time, if $\epsilon^{u}\neq 0$ then $\epsilon^{d}=0$, and vice versa. 
Since we consider flavor-universal diagonal NSI couplings, we do not anticipate modifications to Mikheyev-Smirnov-Wolfenstein~\cite{Wolfenstein:1977ue, Mikheev:1986if} flavor conversion probabilities inside CCSN, contrary to studies focused on a single flavor NSI coupling~\cite{Stapleford:2016jgz, Jana:2024lfm}.
In addition, constraints from existing terrestrial oscillation experiments 
do not directly apply.

\prlsection{Neutral current processes}{.}
\label{sec:NSI-CEvNS} 
In the following we focus on NC neutrino interactions and examine the impact of NSI on both coherent elastic neutrino-nucleus scattering (CE$\nu$NS) and elastic neutrino-proton scattering. The differential CE$\nu$NS cross section for $X(A,Z)+\nu \rightarrow X(A,Z)+\nu$ in the SM is~\cite{Freedman:1973yd}
\begin{equation} 
\label{eq:cevnsxsec}
    \frac{d\sigma_{\nu X}}{dE_r} \simeq \frac{G_\mathrm{F}^2 m_X}{4\pi} Q_W^2\left(1- \frac{m_X E_r}{2E_{\nu}^2}\right)F^2(Q) \ ,
\end{equation}
where $E_\nu$ is the incoming neutrino energy, heavy nucleus $X(A,Z)$ mass is $m_X \simeq 931.5 A\;$MeV with $A$, and $Z$ being respectively the nucleus mass and atomic numbers, $Q_W = [N - Z(1 - 4\sin^2(\theta_W))]$ is the SM weak charge, $\sin^2\theta_W = 0.23863$ is the weak mixing angle at low momentum transfer~\cite{ParticleDataGroup:2022pth}, $N$ is the neutron number and $E_r$ is the nuclear recoil energy.
For low momentum transfer the scattering is fully coherent with $F(Q)\simeq1$, but at higher $Q = \sqrt{2 m_X E_r}$ finite-size effects need to be included; to account for them, we adopt the Helm form factor~\cite{Helm:1956zz}.

The presence of NSI modifies the CE$\nu$NS cross section through the weak charge, which for NSI takes the form
$Q_W^{\rm NSI} = -2~[Z(2\epsilon^u+\epsilon^d) + N(\epsilon^u+2\epsilon^d)]$. This gives an
effective charge of $Q'_W = Q_W + Q_W^\mathrm{NSI}$. The corresponding cross section $\sigma^\mathrm{NSI}_{\nu X}$ is then obtained by replacing $Q_W$ with $Q_W^\prime$ in Eq.~\eqref{eq:cevnsxsec}. The ratio of the NSI-modified and SM cross sections is then $\sigma^\mathrm{NSI}_\mathrm{\nu X} / \sigma^\mathrm{SM}_{\nu X} = Q_W^{'2}/Q_W^2$.

The SM neutrino-proton elastic scattering $\nu$-p cross section for SN neutrino energies is given by~\cite{Beacom:2002hs, Dasgupta:2011wg}
\begin{equation}
\label{eq:proton-scattering}
\frac{d\sigma_{\nu \mathrm{p}}}{dE_\mathrm{p}} \simeq  \frac{G_\mathrm{F}^2 m_\mathrm{p}}{2\pi E_\nu^2} \left( B E_\nu^2 +  C (E_\nu - E_\mathrm{p})^2  -  D m_\mathrm{p} E_\mathrm{p}  \right) \ ,
\end{equation}
where $E_\mathrm{p}$ is the proton recoil energy, $m_\mathrm{p}$ the proton mass, and the coefficients are $B = (c_V + c_A)^2$, $C = (c_V - c_A)^2$, and $D = (c_V^2 - c_A^2)$ with $c_V \simeq 0.02$, $c_A = 1.27/2$ for neutrinos and $c_A = -1.27/2$ for antineutrinos. The modification to the SM $\sigma_{\nu \mathrm{p}}$ cross section due to vector NSI can be accounted for by replacing $c_V$ with 
$c_V^\prime = c_V + 2\epsilon^u + \epsilon^d$. Unlike CE$\nu$NS, however, the ratio of the NSI-modified and the unmodified SM cross sections for $\nu$-p scattering is energy dependent.

Additional NC channels can also complement our analysis. In water Cherenkov detectors, inelastic neutrino scattering on $^{16}$O produces de-excitation $\gamma$-rays below 10 MeV and has been proposed as a probe of SN $\nu_\mu$ and $\nu_\tau$ fluxes~\cite{Langanke:1995he}. Although this signal is obscured by the inverse beta decay background, neutron tagging with gadolinium in Super-Kamiokande~\cite{Super-Kamiokande:2021the} can facilitate post-processing analyses to identify it. Further, potentially promising NC channels from $^{40}$Ar targets in the DUNE liquid argon time projection chamber detector \cite{DUNE:2020zfm} have been put forth~\cite{Tornow:2022kmo, Newmark:2023vup}.
 
\begin{figure}[t]
\centering
\includegraphics[width=.43\textwidth]{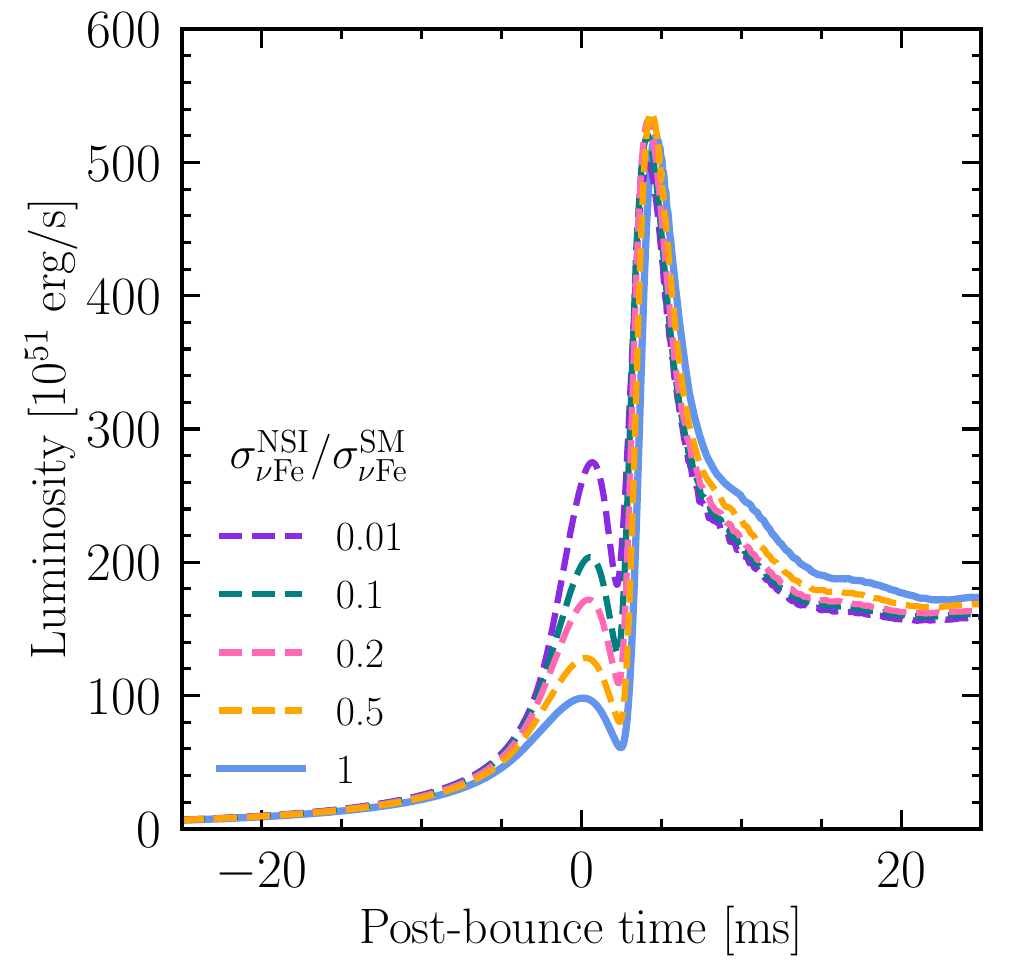}
\caption{All-flavor neutrino luminosity in the observer frame as a function of post-bounce time from our one-dimensional \texttt{GR1D} CCSN simulations with (dashed) and without (solid blue) NSI. The NSI effects correspond to $\epsilon^u<0.16$, where there is a negative interference with the SM contribution for CE$\nu$NS, while the solid line represents the SM case with $\epsilon^u=0$ and $\sigma^\mathrm{NSI}_{\nu\mathrm{Fe}}/\sigma^\mathrm{SM}_{\nu\mathrm{Fe}}=1$.}
\label{fig:luminosity}
\end{figure}

\prlsection{Supernova neutrino emission}{.}
\label{sec:SN-nu}
The standard scenario of the CCSN begins with the infall phase, when the inner core approaches the Chandrasekhar mass and electron degeneracy pressure can no longer provide support against gravity. Rapid collapse leads to enhanced electron neutrino emission primarily from electron captures on nuclei. Once the density reaches approximately a few times $10^{11}~\mathrm{g}~\mathrm{cm}^{-3}$, the neutrino diffusion timescale exceeds the collapse timescale, and neutrinos become trapped through frequent direction-changing scattering off nuclei. This results in a distinct pre-neutronization peak in the neutrino emission signal. 

\begin{table*}[t]
\centering
\caption{
\label{tab:detector_table}
Number of expected detected events from Betelgeuse CCSN at $D = 0.2$ kpc in the time windows used in the statistical analysis for NSI and for SM in the characteristic target detectors considered. We consider one coupling at a time with $\epsilon^u=0.158$ ($\epsilon^d=0$) or $\epsilon^d=0.150$ ($\epsilon^u=0$) resulting in the ratio for CE$\nu$NS $\sigma^\mathrm{NSI}_{\nu \mathrm{Fe}} / \sigma^\mathrm{SM}_{\nu \mathrm{Fe}} = 0.01$. The effective target masses and energy ranges of each detector are also shown. For the calculation of the event rates in the xenon-based detectors we include the energy-dependent efficiency from XENON1T~\cite{XENON:2018voc}. The cross section ratios refer to CE$\nu$NS for the DM detectors (target materials xenon, lead, and argon) and $\nu$-p elastic scattering for JUNO (target material protons).  We also list the assumed $1\sigma$ uncertainties in the SM cross sections (Uncertainty SM X-sec) and CCSN neutrino flux (Uncertainty SN Flux), see Supplemental Material Sec.~\ref{app:C} for further discussion of the uncertainties. Due to the neutrino energy dependence of the ratio of the NSI-modified and the unmodified SM cross sections for $\nu$-p scattering, we quote an upper projection bound value in the ratio column representing the largest expected value for neutrino energies expected from CCSN. For example for $E_\nu=15\;\mathrm{MeV}$ this is approximately 1.06 ($\epsilon^u\neq 0$) and 1.02 ($\epsilon^d\neq 0$).\\}
\begin{tabular}{|c|c|c|c|c|c|c|c|c|c|l}
\hline
 Detector & Target & Energy   & {Uncertainty} & {Uncertainty} & $\sigma^\mathrm{NSI}/\sigma^\mathrm{SM}$ & $\sigma^\mathrm{NSI}/\sigma^\mathrm{SM}$ &  & $N_\mathrm{det}^{\mathrm{NSI}}$ & $N_\mathrm{det}^{\mathrm{NSI}}$\\ 
 (target) & mass & range & {SM X-sec (\%)} & {SN flux (\%)} & $[\epsilon^u=0.158]$& $[\epsilon^d=0.150]$ & {$N_\mathrm{det}^{\mathrm{SM}}$} & $[\epsilon^u=0.158]$ & $[\epsilon^d=0.150]$\\
  & (tons) & (keV) &  &  & &  & {} &  & \\
 \hline \hline
         DARWIN (Xe) &  40 & 1-50 & {10} & {15} & 0.047 & 0.027 & 151 & 8.8 & 5.06\\ \hline 
         RES-NOVA (Pb) &  465 & 1-50 &  {10} & {15} & 0.063 & 0.033 & 22731.6 & 2539.8 & 1317.9\\ \hline 
         ARGO (Ar) &  300 & 25-1,000 &  {10} & {15} & 0.018 & 0.015 & 104.4 & 1.07 & 0.87 \\ \hline
         JUNO (p) & 20,000& 200-10,000 &  {10} & {15} &  $\lesssim$ 1.15 & $\lesssim$ 1.05 & 1802 & 8354 & 8685 \\
         \hline
\end{tabular}
\end{table*}

To analyze the impact of NSI on SN neutrino emission, we model the neutrino signal using the 1D CCSN simulation code~\texttt{GR1D}~\cite{OConnor:2009iuz, OConnor:2014sgn} together with \texttt{NuLib} code \cite{Sullivan:2015kva} for neutrino opacities. For both the SM and NSI scenarios we include the dominant neutrino-nucleus processes, which are electron captures on nuclei and nucleons, and CE$\nu$NS. We adopt a $15~M_\odot$ progenitor with solar metallicity~\cite{Woosley:1995ip} as our benchmark model.
We use the output luminosity of non-electron neutrinos from the simulation as the sum of all four species $\nu_\mu$, $\nu_\tau$, $\bar\nu_\mu$, and $\bar\nu_\tau$.

Modifications from NSI to the CE$\nu$NS cross section can significantly alter the shape of the CCSN pre-neutronization neutrino peak~\cite{Huang:2021enl}. We focus on NSI mediated by heavy particles with masses above a few hundred MeV. Hence, the on-shell production of such particles in the core is negligible since the temperature during infall reaches only $\sim$15 MeV.

In Fig.~\ref{fig:luminosity} we show the luminosities of all neutrino flavors from CCSN simulations with and without NSI effects. During the pre-neutronization burst the luminosity increases in the NSI cases compared to the SM case. The luminosity increase originates from a significantly smaller CE$\nu$NS cross section when NSI are present and the fact that heavy nuclei dominate the matter composition of the SN core during that phase (see also Fig.~\ref{fig:mean_A_SN} in the Supplemental Material).
After core bounce, the luminosities in both cases become nearly identical because nucleons dominate the matter composition and NSI have a significantly smaller impact on their cross sections than for heavy nuclei (see also Fig.~\ref{fig:CS_ratio} in the Supplemental Material). By contrast, variations in electron capture rates in hot SN matter can modify to a similar degree both the pre-burst and neutronization peaks by $\sim 20$-$30\%$~\cite{Sullivan:2015kva}. Therefore, this uncertainty is not crucial in our case, as we expect a very different degree of change in pre-burst and neutronization peaks.
These considerations highlight the need for multidetector, multi-channel strategies to robustly identify NSI signatures.

\begin{figure*}[t]
\centering
\includegraphics[width=.43\textwidth]{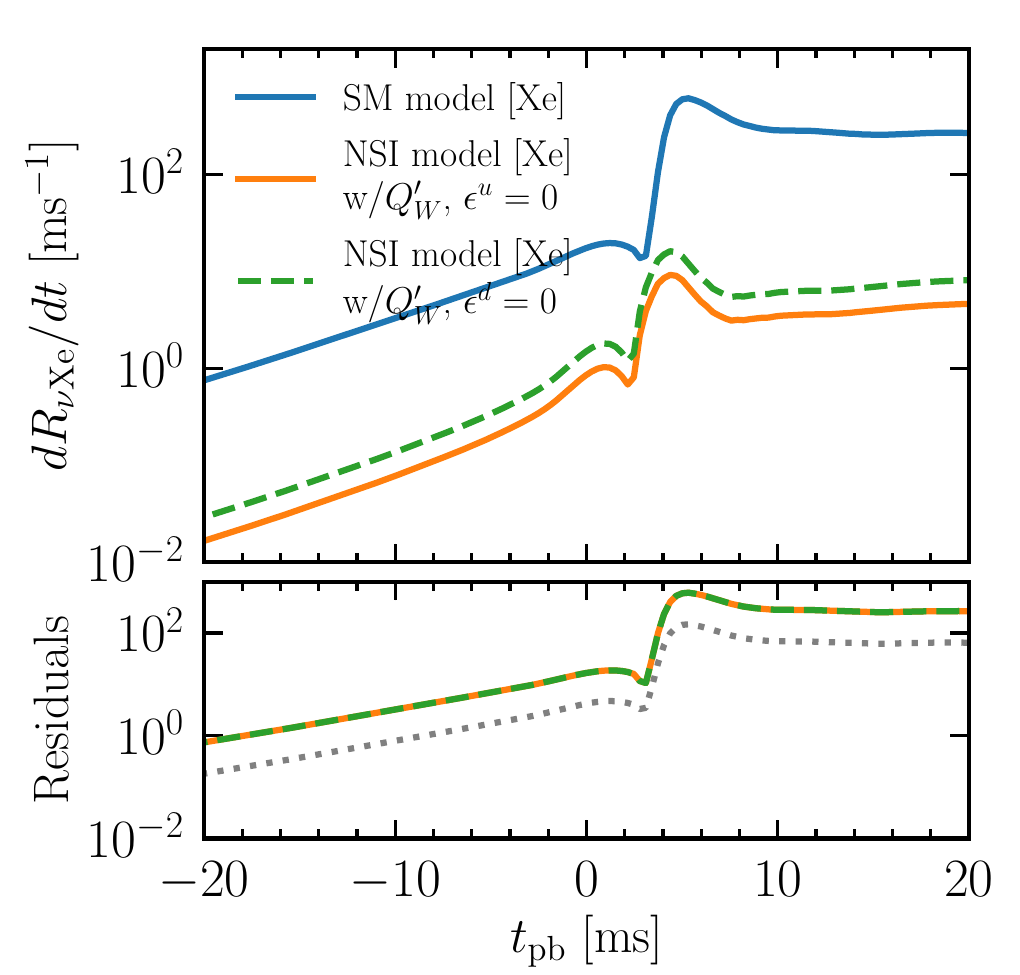}
\includegraphics[width=.43\textwidth]{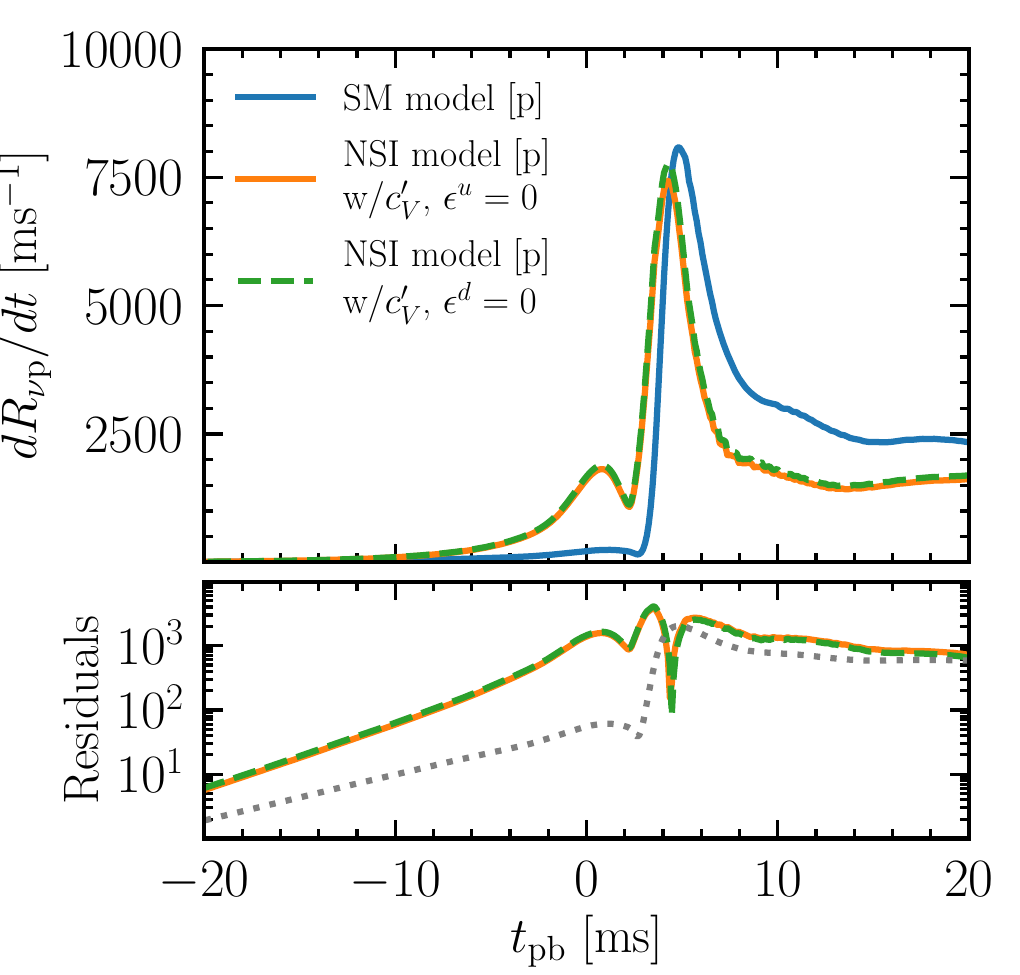}
\caption{
{\it Top panels:} 
Signal event rates for a Betelgeuse CCSN at $D=0.2$ kpc in a xenon-based CE$\nu$NS detector such as DARWIN (left) and a liquid-scintillator detector such as JUNO (right). The SM prediction is shown in blue, while NSI benchmarks with one coupling active at a time are shown in orange  $\epsilon^d=0.150$ ($\epsilon^u=0$) and green $\epsilon^u=0.158$ ($\epsilon^d=0$), as listed in Table~\ref{tab:detector_table}.
{\it Bottom panels:} Residuals of the NSI event rates relative to the SM predictions. The gray dotted line indicates a 25\% uncertainty in the SM rates.}
\label{fig:dRdt}
\end{figure*}

To obtain the time-dependent neutrino flux
 we consider the post-bounce time ($t_{\rm pb}$)
luminosities $L_{\nu_\beta}(t_{\rm pb})$,
mean energies $\langle E_{\nu_\beta} \rangle$ and mean-squared energies $\langle E_{\nu_{\beta}}(t_{pb})^2\rangle$ for each flavor $\beta = \{ \nu_e, \bar\nu_e, 4\nu_x \}$, with $\nu_x = \{ \nu_\mu, \bar\nu_\mu, \nu_\tau, \bar\nu_\tau\}$. The flavor-summed flux from CCSN at distance $D$ is then
\begin{equation}
\label{eq:total_flux}
    \frac{d\psi(E_{\nu})}{dt_\mathrm{pb}} = \sum_{\beta} \frac{L_{\nu_\beta} (t_\mathrm{pb})}{4\pi D^2} \frac{\phi_{\nu_\beta} (E_\nu, t_\mathrm{pb})}{\langle E_{\nu_\beta}(t_\mathrm{pb}) \rangle} \  ,
\end{equation}
where $E_\nu$ is the neutrino energy.
The spectral shape is approximated by a pinched distribution~\cite{Keil:2002in}, where we adopt the pinching parameter $\alpha_\beta\equiv 2.3$ corresponding to a Fermi-Dirac distribution.
We use these fluxes to compute signal event rates in experiments.

\prlsection{Experimental signatures}{.}
\label{sec:Neutrino_Event_Rates}
To disentangle the effects of NSI from those of neutrino flavor conversion in the CCSN core \cite{Duan:2010bg, Tamborra:2020cul, Volpe:2023met} we focus on NC detection channels that are to good approximation equally sensitive to all flavors. We consider signatures in both CE$\nu$NS and elastic $\nu$-p scattering.
For CE$\nu$NS, large scale dark matter (DM) detectors are especially promising owing to their low thresholds, scalable volumes and heavy nuclei. The detection of CCSN neutrinos in such experiments has been extensively studied~\cite{XMASS:2016cmy, Lang:2016zhv, Khaitan:2018wnf, Kozynets:2018dfo}, including sensitivity to pre-SN neutrinos~\cite{Raj:2019wpy}.
Low mediator mass NSI SN detection in large xenon-based DM and lead-based detectors was studied in~\cite{Suliga:2020jfa}. 
In contrast, $\nu$-p elastic scattering can be efficiently observed in liquid scintillator detectors \cite{Beacom:2002hs, Dasgupta:2011wg}, which combine sizable volumes with low energy thresholds and excellent energy resolution.

For our analysis we consider a variety of experimental detector configurations, as described in Table~\ref{tab:detector_table}. 
These follow proposals for near future experiments based on xenon such as DARWIN~\cite{DARWIN:2016hyl}/XLZD~\cite{XLZD:2024nsu}, based on argon such as ARGO~\cite{DarkSide-20k:2017zyg} as well as based on lead such as RES-NOVA~\cite{Pattavina:2020cqc, RES-NOVAGroupofInterest:2022glt}. We consider JUNO~\cite{JUNO:2015zny, Lu:2016ipr}, that is already operating, as a representative carbon-based liquid scintillator detector.

The expected event rate in CE$\nu$NS or $\nu$-p elastic scattering for a given detector ($\mathrm{det}$) is
\begin{align}
\label{eq:Rate_time}
    \frac{dR_\mathrm{det}} {dt_\mathrm{pb}dE_r} =&~ N_T ~ \varepsilon(E_r)  \int_{E_{\nu}^\mathrm{min}}^{E_{\nu}^\mathrm{max}} dE_\nu \frac{d\sigma_{\nu (X,p)}}{dE_r} \\ \nonumber 
    &\times  \frac{d\psi(E_{\nu})}{dt_\mathrm{pb}}  ~ \Theta(E_r^\mathrm{max}-E_r) \ ,
\end{align}
where $N_T$ is the number of interaction targets ($X$ nuclei type for CE$\nu$NS or protons in the case of elastic $\nu$-p scattering), $\varepsilon(E_r)$ is the detection efficiency  and $\Theta$ is the Heaviside step function. The average energies of the CCSN neutrinos are between approximately 10-18~MeV, therefore ${E_{\nu}^\mathrm{min}}$ is the minimal neutrino energy to cause the recoil above observable threshold in the corresponding detector, and ${E_{\nu}^\mathrm{max}}=75\;$MeV. The maximum recoil energy is $E_r^\mathrm{max}=2E_\nu^2/(m_T+2E_\nu)$, with $m_T=m_X$ for CE$\nu$NS and $m_T=m_p$ for $\nu$-p scattering. For $\nu$-p scattering we also include the quenching of the proton recoil energy due to scattering in the scintillator medium, following Refs.~\cite{Dasgupta:2011wg, JUNO:2015zny, Chauhan:2021fzu}. On the timescales relevant to the $\mathcal{O}(10)$~ms burst signal duration typical backgrounds are negligible.

\prlsection{Correlated signal detection}{.}
Figure~\ref{fig:dRdt} shows the time-dependent event rates for a Betelgeuse CCSN at $D=0.2$ kpc in a xenon-based DARWIN detector (left) and the JUNO liquid scintillator detector (right). The upper panels compare the SM prediction (blue) with two NSI benchmarks in which only one coupling is switched on at a time: $\epsilon^u=0.158$ ($\epsilon^d=0$) (green, dashed) or $\epsilon^d=0.150$ ($\epsilon^u=0$) (orange, solid).
Although these benchmark couplings lie outside currently allowed regions, they are useful for illustrating the characteristic NSI imprint on the event rates. Table~\ref{tab:detector_table} lists the NSI couplings employed in Fig.~\ref{fig:dRdt}, considered signal energy ranges, and assumed detector specifications. For CE$\nu$NS the suppression factor $\sigma^{\rm NSI}/\sigma^{\rm SM}$ is fixed by the chosen $\epsilon$ and nuclear target, whereas for $\nu$-p scattering the corresponding ratio retains energy dependence. The lower panels show residuals of NSI rates relative to the SM predictions, calculated as the absolute difference between the SM and NSI rates.

In DARWIN the NSI event rate is suppressed compared to SM prediction. Although the neutrino luminosity during the pre-neutronization burst is nearly 4 times larger the CE$\nu$NS cross section for xenon is suppressed by about a factor of 22. For the other CE$\nu$NS detectors considered, ARGO and RES-NOVA, the NSI impact is similar to DARWIN, with consistently reduced rates (see Fig.~\ref{fig:ARGO_RES-NOVA} in the Supplemental Material).  In contrast, in JUNO the NSI event rate during the burst exceeds the SM case, reflecting a small positive modification to the $\nu$-p cross section.  

We evaluate the statistical significance with which a future detection of supernova neutrinos in JUNO and a large-scale DM detector could constrain the NSI parameter space using $\Delta\chi^2$ test. Our analysis includes uncertainties as nuisance parameters for deriving sensitivities. For the details see the Supplemental Material Sec.~\ref{app:C}  

\begin{figure*}[t]
\centering
\includegraphics[width=.42\textwidth]{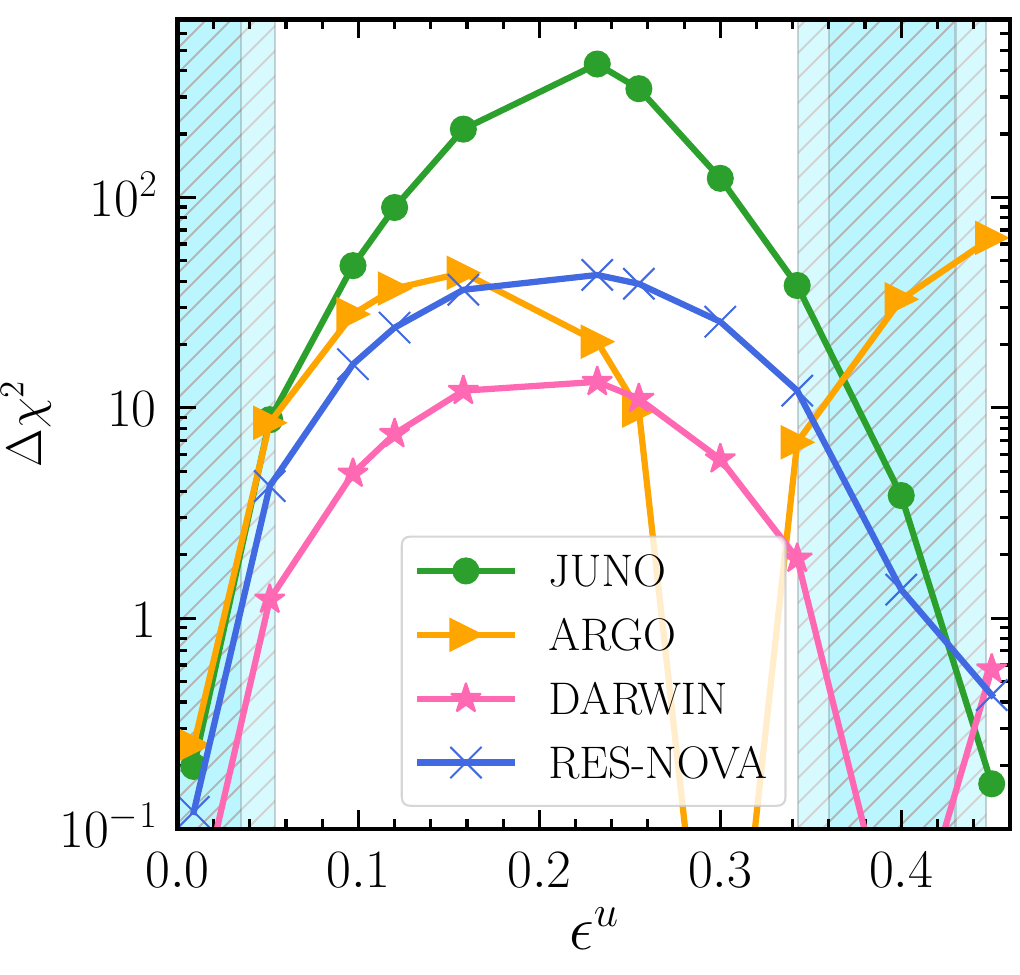}
\includegraphics[width=.42\textwidth]{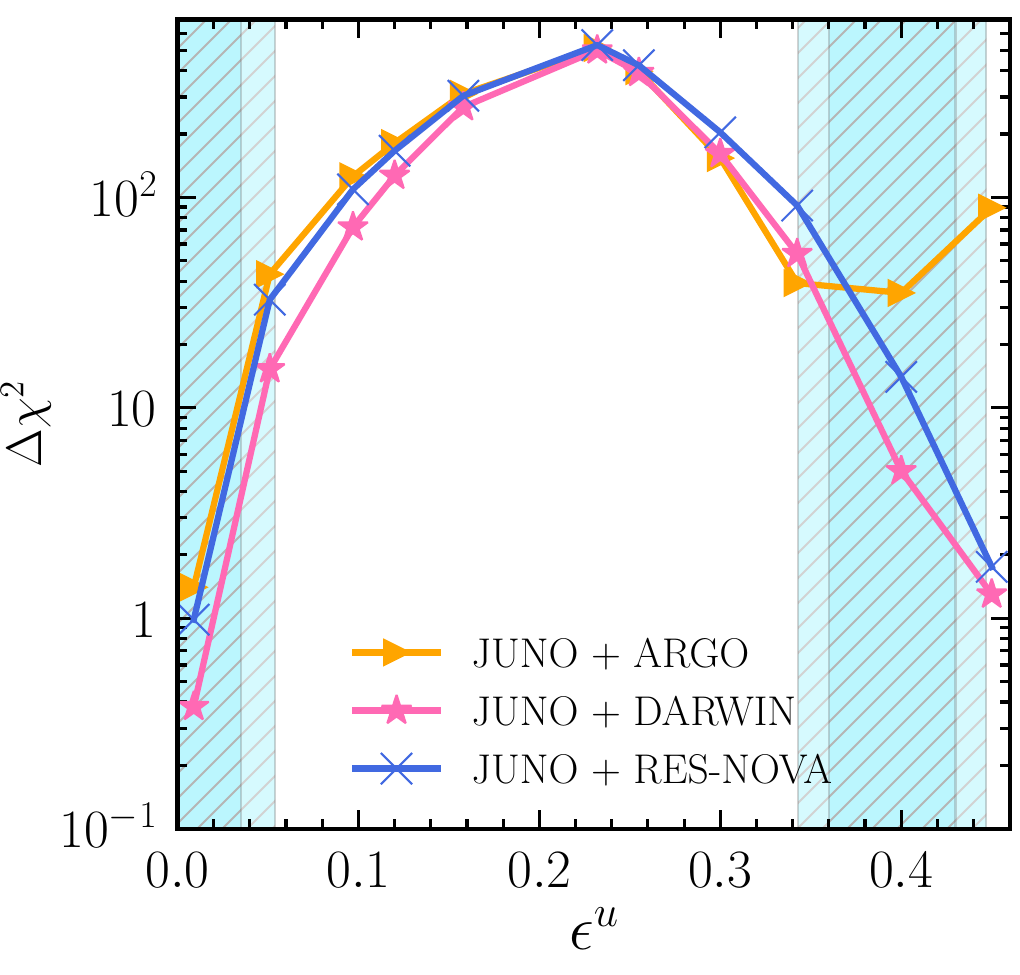}
\caption{{\it Left panel:} Statistical significance ($\Delta\chi^2$) for constraining neutrino-quark NSI as a function of $\epsilon^u$ with individual detectors such as JUNO (green circles), DARWIN (pink stars), ARGO (orange triangles) and RES-NOVA (blue crosses),
assuming Betelgeuse supernova at $D=0.2$ kpc. {\it Right panel:} Combined $\Delta\chi^2$ for detector pairs JUNO+ARGO (orange triangles), JUNO+DARWIN (pink stars), and JUNO+RES-NOVA (blue crosses). In all cases we adopt a supernova flux normalization uncertainty $\sigma_x=0.15$, $\nu$-p cross section uncertainty $\sigma_y=0.1$, and CE$\nu$NS cross section uncertainty $\sigma_z=0.1$. The vertical blue bands mark the 2$\sigma$ (darker) and 3$\sigma$ (lighter) allowed regions for $\epsilon^u_{\mu\mu}=\epsilon^u_{ee}$ from COHERENT CsI+LAr data adopted from Ref.~\cite{Coloma:2022avw}.}
\label{fig:chi-2}
\end{figure*}

The left panel of Fig.~\ref{fig:chi-2} shows the individual detector $\Delta\chi^2$ values for JUNO, ARGO, DARWIN and RES-NOVA as a function of $\epsilon^u$. We also display allowed regions at 2$\sigma$ and 3$\sigma$ from COHERENT CsI+LAr data~\cite{COHERENT:2017ipa, COHERENT:2020iec} adopted from~Ref.~\cite{Coloma:2022avw} (see also e.g.~\cite{DeRomeri:2022twg, Liao:2024qoe}). 
The panel also illustrates for which values of $\epsilon^u$ the detectors with different target materials are individually not sensitive to NSI due to the degeneracy of the SM CE$\nu$NS cross sections. Since for a given value of the $\epsilon$ the cross sections for various target materials are different (see Fig.~\ref{fig:CS_ratio} in the Supplemental Material), and the energy thresholds of various detectors are also distinct, the sensitivity of each of these detectors for a given NSI $\epsilon$ value is different.

We explicitly demonstrate in the
right panel of Fig.~\ref{fig:chi-2} that 
multidetector coincidence detection in
JUNO with ARGO, JUNO with DARWIN and JUNO with RES-NOVA are advantageous compared to individual detector observations.
This reflects the anti-correlation discussed earlier. While NSI increase the signal rates in JUNO, they decrease signal rates in CE$\nu$NS DM detectors for most of the considered $\epsilon^u$ values.
Our sensitivity projections for detected CCSN neutrinos from Betelgeuse allow independently probing the parameter space of terrestrial-based COHERENT CsI+LAr limits.
Notably, combinations of JUNO+ARGO and JUNO+RES-NOVA
allow exclusion of the region at more than the $3\sigma$-level where COHERENT sensitivity is restricted and degenerate around $\epsilon^u=0.4$.
We have verified that increasing the distance to CCSN to 1$~$kpc and increasing the assumed uncertainties do not significantly affect our findings (see Fig.~\ref{fig:chi-2-low-unc} in the Supplemental Material).

Our analysis is complementary to other NSI probes and DM detectors have also been identified as excellent targets for probing NSI with solar neutrinos~\cite{Harnik:2012ni, Cerdeno:2016sfi, Boehm:2018sux, AristizabalSierra:2019ykk, Suliga:2020jfa, Schwemberger:2022fjl, Schwemberger:2023hee, Amaral:2023tbs}. Recent $>2.5\sigma$ observations of solar neutrinos by XENONnT~\cite{XENON:2024ijk} and PandaX-4T \cite{PandaX:2024muv} also allow constraining NSI couplings~\cite{AristizabalSierra:2024nwf, Li:2024iij, DeRomeri:2024iaw, Blanco-Mas:2024ale, Maity:2024aji, Gehrlein:2025isp} at a level comparable to the COHERENT CsI+LAr bounds already shown in Fig.~\ref{fig:chi-2}.
Future CE$\nu$NS detectors should also achieve superior sensitivity to the NSI (see, e.g., Refs.~\cite{COHERENT:2022nrm, AtzoriCorona:2025ygn, AtzoriCorona:2025ibl}). Our results would then serve as a complementary probe.

In the context of SM effective field theory NSI analysis such as Ref.~\cite{Coloma:2024ict}, which uses flavor-preserving NSI constraints from neutrino oscillation experiments, or Ref.~\cite{Breso-Pla:2023tnz}, which combines COHERENT constraints with other electroweak measurements, the degenerate region of the parameter space near $\epsilon\sim$ 0.4 is generally disfavored. However, it is worthwhile to note that these limits assume either a single active flavor coupling or a clear distinction between couplings of different flavors or charged-current NSI acting at the same time or couplings to charged leptons. In the case of a neutral-current democratic NSI neutrino-quark coupling scenario explored in our work, the applicability of these bounds is consequently limited.

We expect that extending our bottom-up approach focused on positive diagonal democratic NSI couplings to negative couplings would lead to less stringent sensitivity limits. In such a case, both the CE$\nu$NS cross section and $\nu$-p cross section would be affected in the same direction, and the gain from using detectors with these interaction channels would be smaller.

\prlsection{Conclusions}{.} 
\label{sec:Conclusions}
Using self-consistent modeling we show neutrino-quark NSI alter pre-neutronization supernova emission yielding oscillation-independent NC imprints. Coincidence detection between liquid scintillator JUNO and DM experiments such as DARWIN/XLZD, ARGO or RES-NOVA exploiting opposing NSI effects on $\nu$-p and CE$\nu$NS event rates allows
independent tests of NSI couplings in the $\epsilon^u \simeq 0.035-0.4$ range potentially beyond terrestrial limits for a Galactic supernova within $\sim1$ kpc, breaking degeneracies with the SM effects plaguing conventional searches. More broadly, our results establish multidetector supernova neutrino observations as precision laboratories for fundamental neutrino interactions complementary to and in some regimes surpassing terrestrial searches. This method is broadly applicable to a wide range of upcoming experiments and introduces a new principle for testing fundamental interactions in extreme astrophysical environments beyond terrestrial searches in degenerate regimes.


\begin{acknowledgments}
\textbf{Acknowledgments.---}
 We thank Luca Boccioli, Peter Denton, Masayuki Nakahata, and Michael Smy for useful discussions.
A.M.S. and A.R.B. acknowledge support from NSF N3AS Physics Frontier Center, NSF Grant No. PHY-2020275, and the Heising-Simons Foundation (2017-228). 
The work of A.M.S. is supported by the Neutrino Theory Network Program Grant No. DE-AC02-07CHI11359. V.T. acknowledges support by the World Premier International Research Center Initiative (WPI), MEXT, Japan and JSPS KAKENHI grant No. 23K13109.
A.M.S. and V.T. also acknowledge a partial support from the Center for Theoretical Underground Physics and Related Areas (CETUP*) and its kind hospitality and stimulating research environment.

\end{acknowledgments}
 
\bibliography{NSI-SN-DET}
\phantom{i}
\clearpage

\appendix


\clearpage
\newpage
\onecolumngrid

\centerline{\large {Supplemental Material for}}
\medskip

{\centerline{\large \bf{Oscillation-independent probes of nonstandard neutrino interactions from
supernovae}}}
\medskip
{\centerline{Angela R.~Beatty, Anna~M.~Suliga, Volodymyr Takhistov}}
\bigskip
\bigskip

In this Supplemental Material we provide additional figures complementing the main text describing effects of NSI on neutrino interaction cross sections as well as detector event rates. Further, we provide additional details of coincident detection statistical analysis.  

We assume a heavy mediator with mass $\gtrsim 100~\mathrm{MeV}$ that induces effective four-fermion NSI with first-generation quarks only. 
The couplings are taken to be diagonal and flavor-universal and we vary one quark coupling at a time ($\varepsilon^u$ or $\varepsilon^d$) for direct comparison to literature.
We focus on NC detection channels, which renders our approach insensitive to flavor conversions.

\section{Additional Figures for Cross Section NSI Effects}

\begin{figure*}[b]
\centering
\includegraphics[width=.43\textwidth]{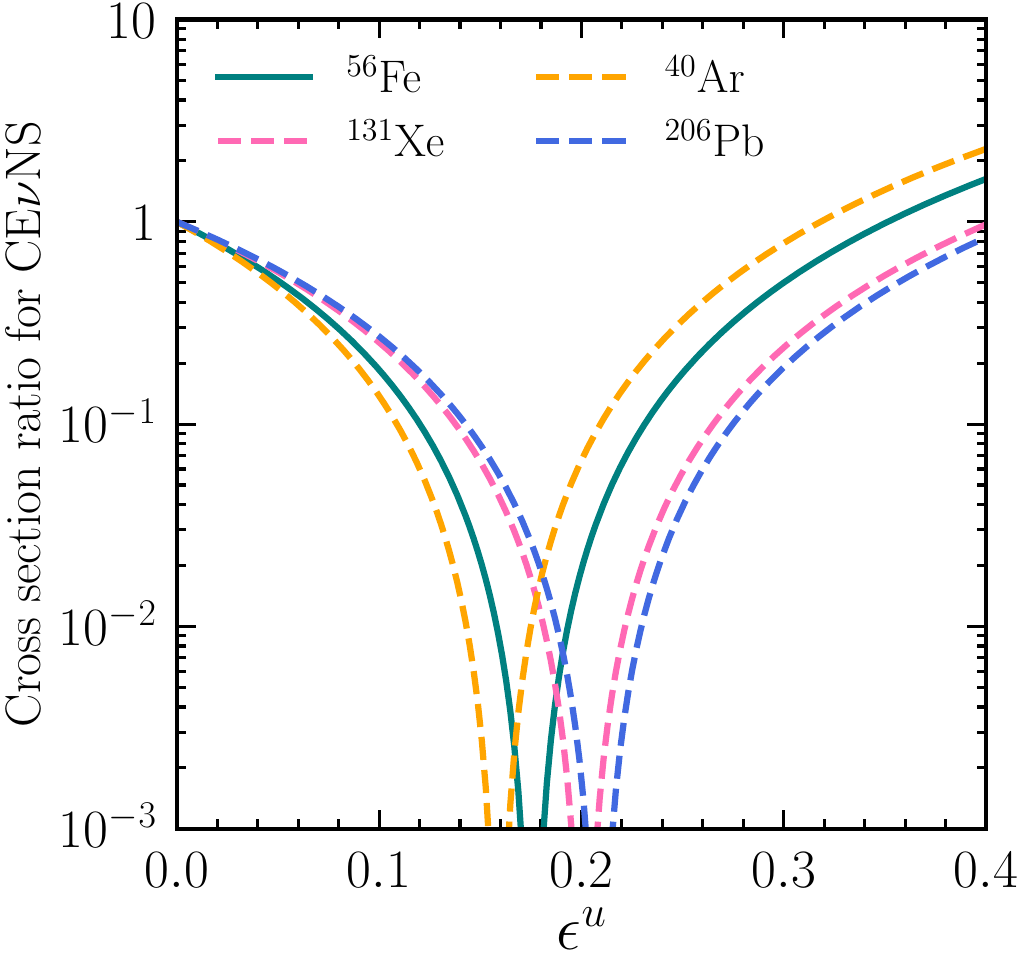}
\includegraphics[width=.43\textwidth]{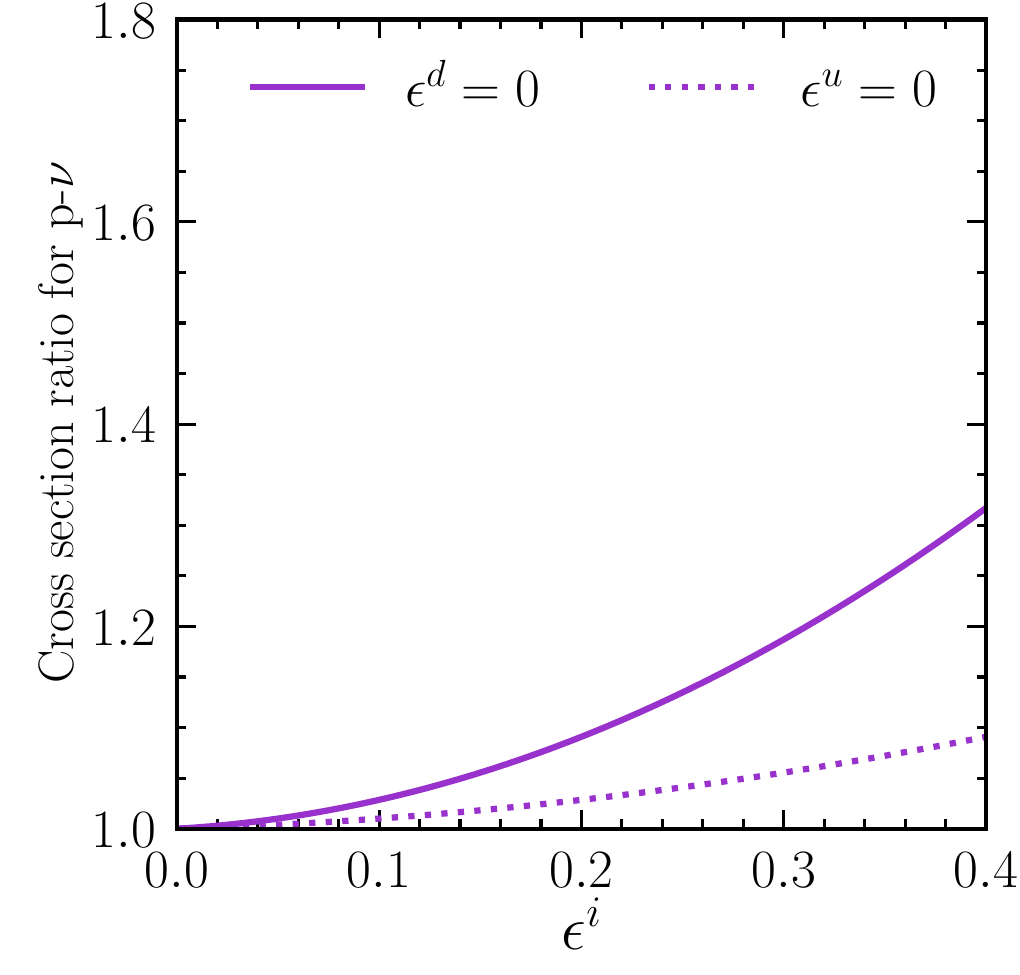}
\caption{Ratios of the NSI cross section to the SM cross section for different elements. \emph{Left panel:} Ratios of the CE$\nu$NS cross sections as a function of $\epsilon^u$ for ${}^{56}$Fe (solid green line),  ${}^{131}$Xe (dashed pink line), ${}^{40}$Ar (dashed yellow line),  ${}^{206}$Pb (dashed blue line). \emph{Right panel:} Ratio of the NC $\nu$-p cross section as a function of $\epsilon^u$ (solid purple line) and $\epsilon^d$ (dotted purple line) for $E_\nu=15\;\mathrm{MeV}$ integrated over the proton recoil energy.}
\label{fig:CS_ratio}
\end{figure*}

Figure~\ref{fig:CS_ratio} shows the ratios of the NSI to SM cross sections for CE$\nu$NS [left panel] for ${}^{56}$Fe (solid green line), ${}^{131}$Xe (dashed pink line), ${}^{40}$Ar (dashed yellow line), ${}^{206}$Pb (dashed blue line), and $\nu$-p scattering [right panel]  (solid and dotted purple lines). Most of the NSI parameter space considered in our work for $\epsilon^u$ results in a negative interference with the SM couplings resulting in the suppression of the CE$\nu$NS cross sections.

\begin{figure*}[t]
\centering
\includegraphics[width=.45\textwidth]{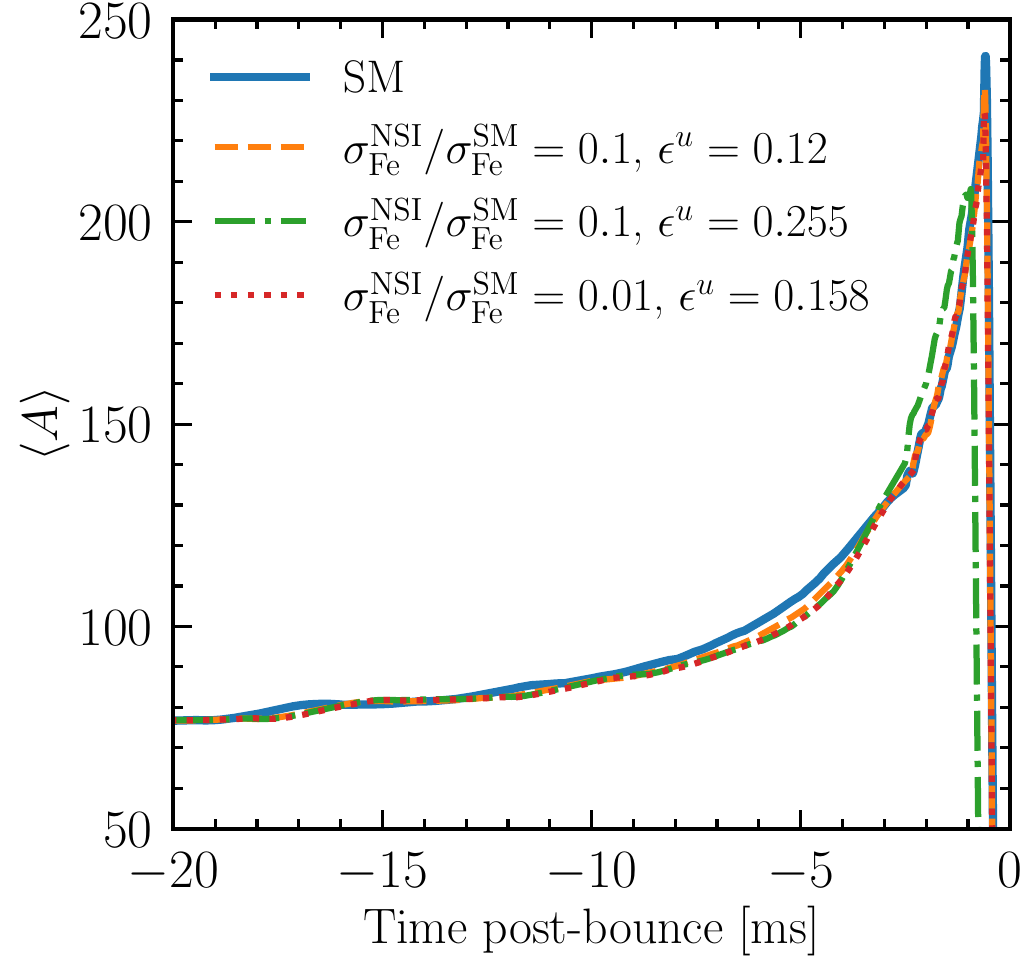}
\includegraphics[width=.45\textwidth]{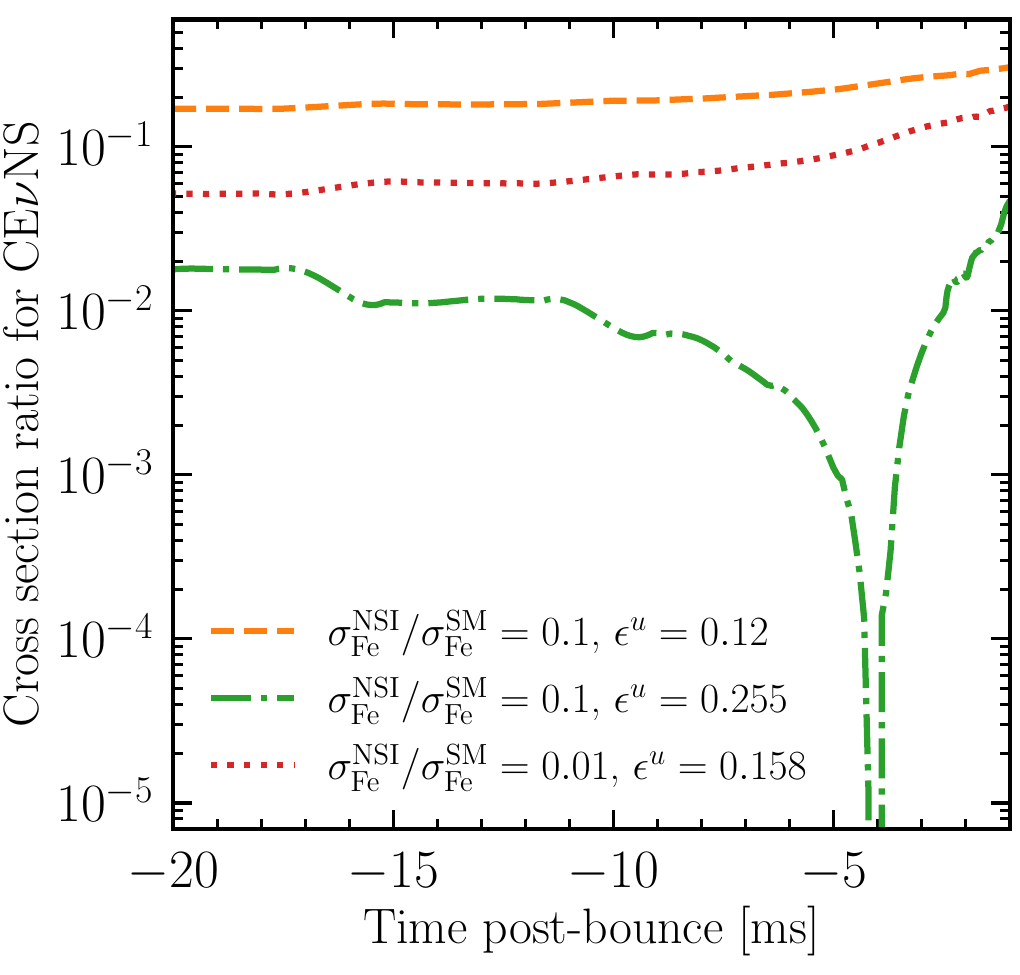}
\caption{\emph{Left panel:} Comparison of the evolution of the mean nuclear mass during the pre-bounce CCSN evolution simulated with the SM CE$\nu$NS cross section (solid line) and the NSI CE$\nu$NS cross sections (dashed lines). \emph{Right panel:} The evolution of the ratios of the NSI to the SM cross section as a function of the post-bounce time for CCSN simulations with different values of $\epsilon^u$.}
\label{fig:mean_A_SN}
\end{figure*}

The left panel of Fig.~\ref{fig:mean_A_SN} illustrates the changes of the temporal evolution of mean nuclear mass $\langle A\rangle$ due to the NSI in the center of the SN core for a set of SN simulations performed using~\texttt{GR1D}~\citep{OConnor:2009iuz, OConnor:2014sgn} with the Lattimer and Swesty equation of state~\citep{Lattimer:1991nc}. The $\langle A\rangle$ tends to decrease for NSI parameters suppressing the neutrino scattering with nuclei. This effect is likely a consequence of an increased deleptonization of the SN core. The right panel of Fig.~\ref{fig:mean_A_SN} shows the temporal evolution of the CE$\nu$NS cross sections ratio inside the supernova core for a set of NSI parameters.  

\section{Additional Figures for Detector Event Rates}
\label{app:B}
\begin{figure*}[h]
    \centering
    \includegraphics[scale=.45]{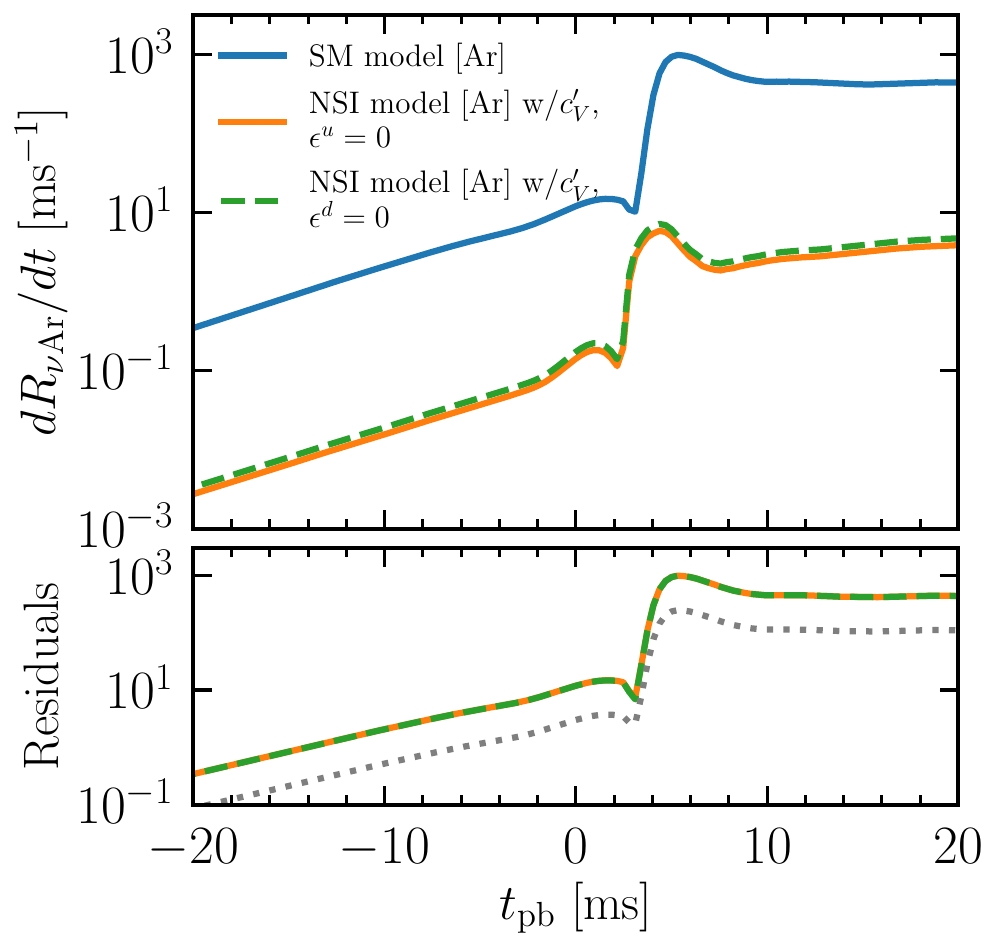}
    \includegraphics[scale=.45]{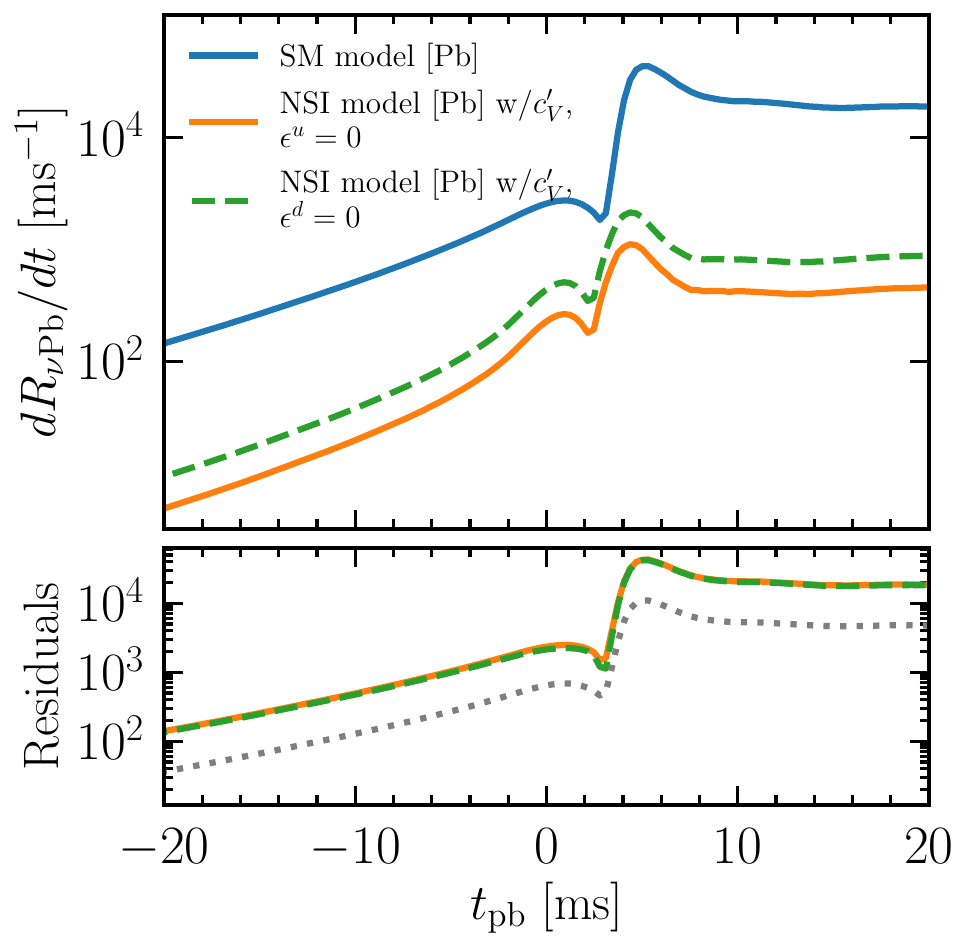} 
    \caption{Same as Fig.~\ref{fig:dRdt} but for argon-based CE$\nu$NS detector like ARGO (left panel) and lead-based detector like RES-NOVA (right panel). The values of $\epsilon^u$ and $\epsilon^d$ used for the event rate calculations are listed in Table~\ref{tab:detector_table}.}
    \label{fig:ARGO_RES-NOVA}
\end{figure*}

Figure~\ref{fig:ARGO_RES-NOVA} shows the SM and NSI modified event rates for Betelgeuse CCSN neutrinos ($D=0.2\;$kpc) in the ARGO (left panels) and RES-NOVA (right panels) together with the residuals of the NSI rates relative to the SM in the corresponding lower panels.  
In the NSI case, the numerical SN simulations include the modifications to the CE$\nu$NS and $\nu$-p scattering with $\epsilon^u=0.158$, resulting in the ratio for CE$\nu$NS $\sigma_\mathrm{NSI} / \sigma_\mathrm{SM} = 0.01$ for $^{56}$Fe. The detector characteristics used to calculate the event rates are listed in Table~\ref{tab:detector_table}.

\section{Details of Coincident Detection Statistical Analysis}
\label{app:C}
We evaluate statistical significance for the NSI discrimination with the supernova neutrino detection using  $\Delta\chi^2$ test statistic with Gaussian $1\sigma$ pull terms \cite{Fogli:2002pt}.
The combined sensitivity for two-detector combination is given by 
\begin{equation}
\label{eq:chi2}
    \Delta \chi^2   = \min_{x,y,z} \left\{  \Delta\chi^2_\mathrm{JUNO} +  \Delta\chi^2_\mathrm{DM}  + \sum_{l=x,y,z}\left(\frac{l}{\sigma_l}\right)^2 \right\}  \, .
\end{equation}
where we assume that $\sigma_x=15\%$ is the supernova neutrino signal uncertainty  during early CCSN signal~\cite{OConnor:2018sti}, $\sigma_y=10\%$ is the NC $\nu$-$\mathrm{p}$ cross section uncertainty, and $\sigma_z=10\%$ is the CE$\nu$NS cross section uncertainty. We minimize the statistic over the free parameters $\{x, y, z\}=(-0.99, 5]$ associated with the corresponding $1\sigma$ uncertainties.

We have chosen the $1\sigma$ uncertainty values in the considered cross sections by considering both the experimental and theoretical uncertainties. In the case of CE$\nu$NS, the CsI experimental uncertainty is at the level of approximately 15\% \cite{COHERENT:2021xmm}, but the theoretical predictions reach percent level uncertainty~\cite{Tomalak:2020zfh}. For the NC $\nu$-p cross section the uncertainty spans from the strange axial contribution $\Delta s$.  While historic experiments allow $\Delta s$ ranges that could shift the cross section by up to $\sim25\%$ for $\Delta s = \pm 0.15$~\cite{Ahrens:1986xe, Chauhan:2022wgj}, recent lattice QCD determinations significantly constrain this $\Delta s = -0.018 \pm 0.006$~\cite{Chambers:2015bka}, implying sub-percent uncertainty for the NC $\nu$-p cross section. Therefore, in our cross section calculation, we assume this contribution to be negligible, and we only include the variation to the cross section through the uncertainty $\sigma_y$.

The single detector $\Delta\chi^2_\mathrm{det}$ terms are
\begin{equation}
   \Delta\chi^2_\mathrm{det} = \frac{\left(N_\mathrm{det}^\mathrm{SM} \left(1+x\right)(1+i) - N_\mathrm{det}^\mathrm{NSI}\right)^2}{N_\mathrm{det}^\mathrm{SM}} \,
\end{equation}
where the total number of events in a single detector $\mathrm{det}=\{\mathrm{JUNO, DARWIN, ARGO, RES}$-$\mathrm{NOVA}\}$, with 
$i=y$ for JUNO and $i=z$ for the DM detectors,
the total number of events 
in the assumed time and energy window is $N^{\mathrm{SM}}_\mathrm{det} = N_\mathrm{det}\left(\epsilon^u=0\right)$ for the SM case and $N^{\mathrm{NSI}}_\mathrm{det} = N_\mathrm{det}\left(\epsilon^u\neq0\right)$ for the NSI modified case, with the case of $\epsilon^d \neq 0$ being analogous. The total time and energy integrated number of events for a given detector is
\begin{equation}
N_\mathrm{det} =  \int_{t_0}^{t_1}dt \int_{E_r^\mathrm{min}}^{E_r^\mathrm{max}} dE_r \frac{dR_\nu}{dt_\mathrm{pb}dE_r}  \ , 
\end{equation}
with integration energy ranges listed in Table~\ref{tab:detector_table}. 
For JUNO, the integration time extends to the dip between the pre-neutronization and neutronization bursts, while for the DM detectors we integrate to 75 ms after bounce. These assumptions yield conservative projections. Comprehensive time-dependent and energy-dependent analyses can be expected to be even more sensitive.

\begin{figure*}[h!]
\centering 
\includegraphics[width=0.45\textwidth]{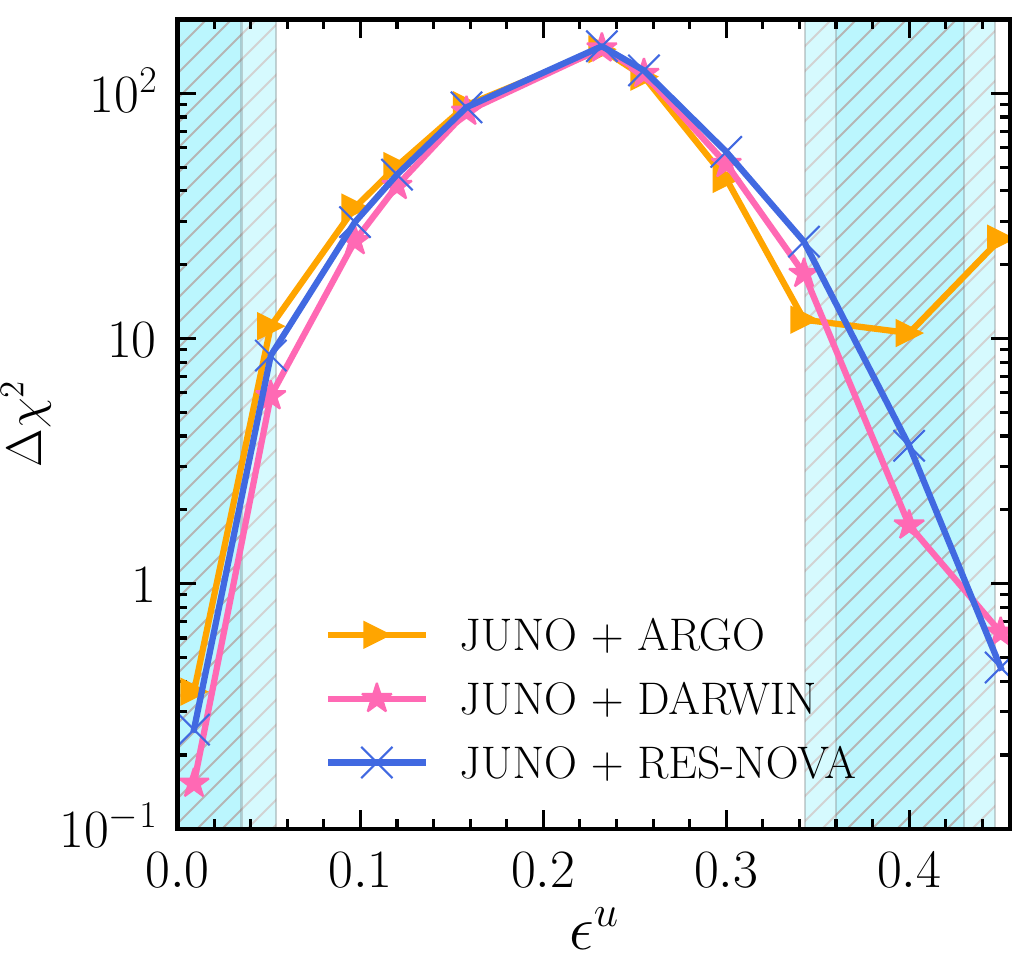}
\includegraphics[width=0.45\textwidth]{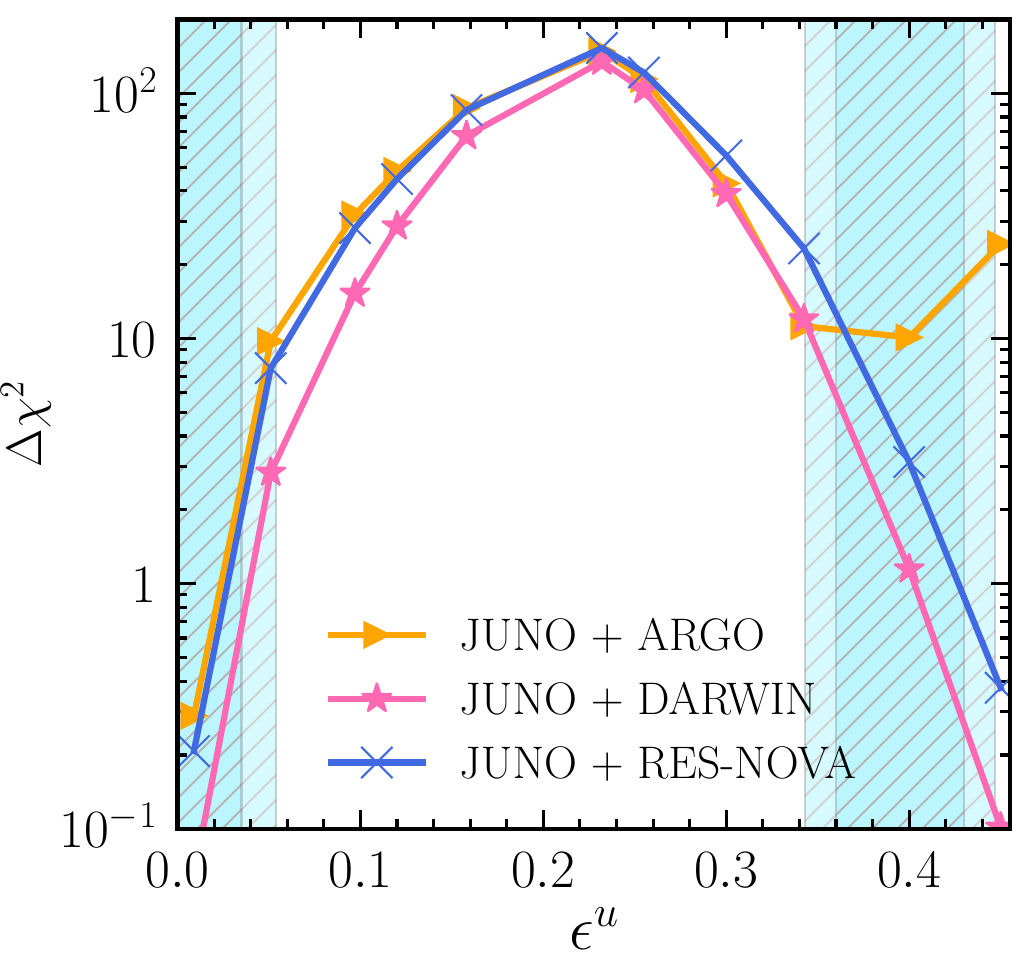}
\caption{\emph{Left Panel:} Same as Fig.~\ref{fig:chi-2}, but with increased uncertainties $\sigma_x=25\%$, $\sigma_y=20\%$, and $\sigma_z=20\%$. \emph{Right Panel:} Same as the left panel but the distance to the supernova $D=1\;$kpc.} 
\label{fig:chi-2-low-unc}
\end{figure*}

Figure~\ref{fig:chi-2-low-unc} depicts the results of our statistical analysis for two different cases with respect to the result presented in the main text: (i) the SN neutrino flux and cross section uncertainties are increased to $\sigma_x=25\%$, $\sigma_y=20\%$, and $\sigma_z=20\%$ with respect to Fig.~\ref{fig:chi-2} (left panel) and (ii) the uncertainties are large and the distance to the SN is increased to $D=1\;$kpc compared to Fig.~\ref{fig:chi-2}. The increase in the uncertainty or distance to the SN does not change our results significantly.

\end{document}